\documentclass[%
 amssymb, amsmath,%
 aip,cha,%
]{revtex4-1}

\usepackage{docs}%
\usepackage{bm}%
\usepackage[colorlinks=true,linkcolor=blue]{hyperref}%
\usepackage[dvips]{graphicx}
\expandafter\ifx\csname package@font\endcsname\relax\else
 \expandafter\expandafter
 \expandafter\usepackage
 \expandafter\expandafter
 \expandafter{\csname package@font\endcsname}%
\fi
\begin{document}

\title{Mode bifurcation on the rythmic motion of  a micro-droplet under stationary DC electric field}

\author{Tomo Kurimura}%
\email{kurimura@chem.scphys.kyoto-u.ac.jp}
\affiliation{Department of Physics, Graduate School of Science, Kyoto University, Kitashirakawa-Oiwake-cho, Sakyo-ku, Kyoto 606-8502 Japan}%

\author{Masahiro Takinoue}%
\affiliation{Interdisciplinary Graduate School of Science and 
Engineering, Tokyo Institute of Technology, 4259 Nagatsuta-cho, Midoriku, Yokohama, Kanagawa 226-8503, Japan}%

\author{Masatoshi Ichikawa}%
\affiliation{Department of Physics, Graduate School of Science, Kyoto University, Kitashirakawa-Oiwake-cho, Sakyo-ku, Kyoto 606-8502 Japan}%

\author{Kenichi Yoshikawa}%
\affiliation{Faculty of Biological and Medical Sciences, Doshisha Univ.1-3 Tataramiyakodani, Kyotanabe, Kyoto 610-0394, Japan}%

\date{May 2012}%

\maketitle


\section{Introduction}
\indent Accompanied by the development of microtechnology, such as MEMS and $\mu$TAS, 
there is increasing interest on the methodology to realize a desired motion of a micro object in a solution environment. 
It is well known that the principle to create an electric motor in a macro system is not applicable to micro system 
because of the enhanced sticky interaction and higher viscosity in micrometer sized system. 
On the other hand, living organisms generate various motions on microscopic scale under isothermal condition. 
Despite the past intensive studies\cite{hiratsuka,vandenheuvel}, the underlying mechanism of the biological molecular motors has not been fully unveiled yet. Under such development status of science and technology on micro-motor sat the present, 
we report a simple motoring system which work smoothly in a microscpic scale. \par
Recently, we found that rhythmic motion is generated for an aqueous droplet in an oil phase under DC voltage on the order of 50 - 100 V.
We have already reported some experiments and models for a w/o droplet under DC electric field.\cite{hase,takinoue}
There are some reports about experiments of w/o droplets moving under electrical field\cite{jung,mochizuki},
bouncing and being absorbed on a surface between water and oil\cite{ristenpart},
deforming and spliting\cite{Eow,teh}.
Manipulating this kind of droplet, which is interesting as the model of the cell\cite{pietrini, tawfik, hase2}, the micro-sized reactor,
by optical tweezers\cite{katsura}, by micro channel\cite{atencia, link}.
And manipulating the cells or micro objects by electrical field has been attempted\cite{voldman},
to know manipulating this kind of droplets in detail will help this in the future.
In the present article, we will show that rhythmic motion on micro-droplet is induced under the DC potential on the order of several volts.
We will also propose a simple mathematical model to reproduce the rhythmic motion and mode bifurcation.

\section{Experimental}

\begin{figure}[htbp]
\begin{center}
\includegraphics[]{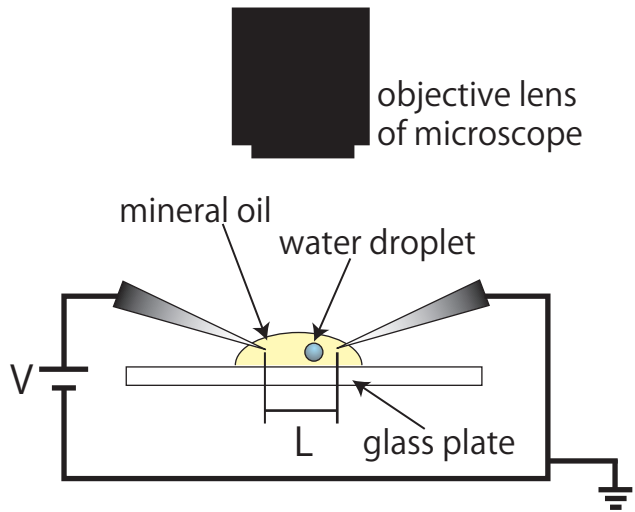}
\end{center}
\caption{Schematic representation of the experimental setup. 
Mineral oil containing water droplets was placed on a glass slide 
and couple of electrodes was situated inside the oil phase. 
V: Applied DC voltage. 
L: distance between the electrodes.\label{fig:one}}
\end{figure}

\indent A schematic illustration of the experimental setup is given in FIG.\ref{fig:one}.
A water droplet was suspended in mineral oil on a glass slide, and
constant voltage was applied to the droplet using cone-shaped tungsten electrodes.
Droplet motion was observed using an optical microscope (KEYENCE 
, Japan).\par

The w/o droplet was generated using a vortex mixer as follows.
We prepared mineral oil including surfactant:
10$\mu$m surfactant, dioleylphosphatidylcholine (DOPC) (
Japan), was
solved in mineral oil (Nacalai Tesque, Japan) by 90 min sonication at $50\ {}^\circ\mathrm{C}$.
2$\mu$l ultrapure water (Millipore, Japan) was added to 100 $\mu$l of the prepared mineral oil, and
then agitated by a vortex mixer for approximately 3 s.

\section{Results}
\begin{figure}[htbp]
\begin{center}
\includegraphics[]{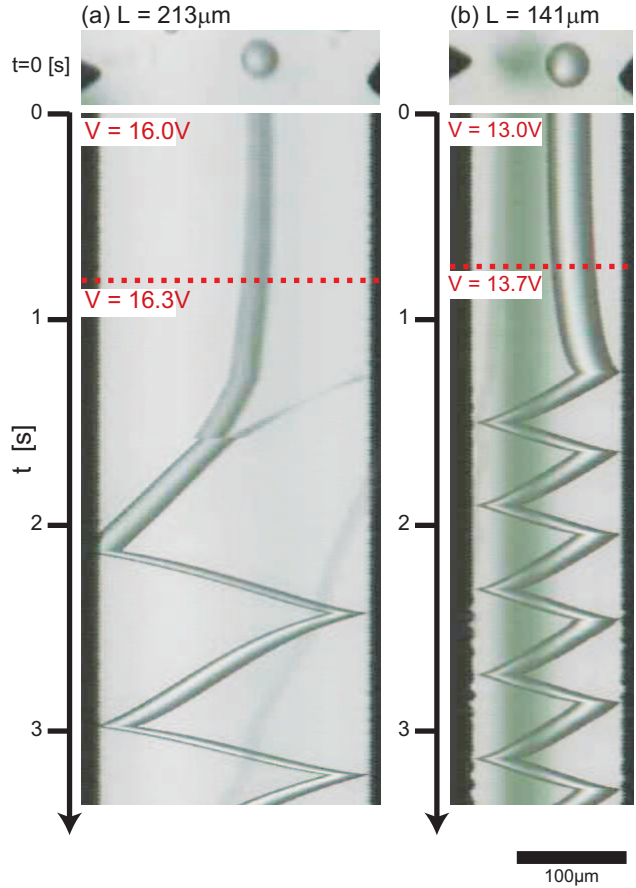}
\end{center}
\caption{Spatio-temporal diagram on the motion of a droplet with the diameter of 34 μm 
at (a) L=213μm, (b) L=141μm.　Bifurcation from the stationary state into an oscillatory state 
is induced by the increase of the applied voltage. \label{fig:two}}
\end{figure}

\indent FIG.\ref{fig:two} exemplifies the motion of a droplet under DC electric field, 
indecating the occurrence of the periodic go-back motion 
between the electrodes accompanied by the increase of the electical potential.  
In the experiments, we observed two following types of behavior: 
oscillatory and stationary.
These behaviors switch each other depending on the applied voltage.
When the distance between two electrodes was 213$\mu$m [FIG.\ref{fig:two}(a)],
the droplet started the motion with the applied voltage above 16.3 V.
When the distance between two electrodes was 141$\mu$m [FIG.\ref{fig:two}(b)],
the droplet started moving with the applied voltage above 13.7 V.\par
\begin{figure}[htbp]
\begin{center}
\includegraphics[]{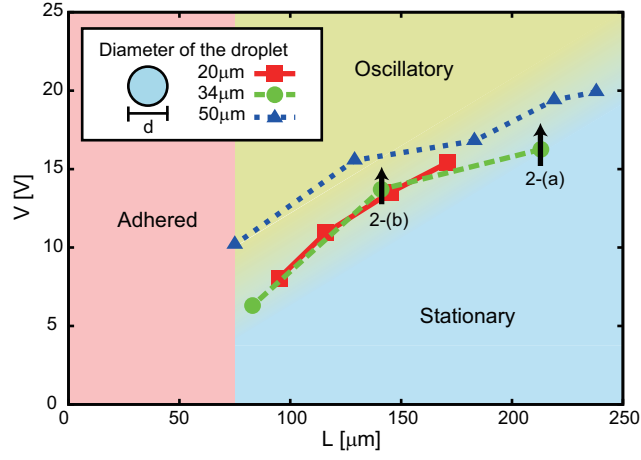}
\end{center}
\caption{Phase diagram for the mode bifurcation between rhythmic motion 
and stationary state as obderved for the droplets with different diameter,
where each point represent the threshold value on the bifurcation. 
The two arrows are correspond to the mode bifurcation as shown in FIG.\ref{fig:two}\label{fig:three}}
\end{figure}

FIG.\ref{fig:three} shows the diagram of the mode of droplet behavior 
depending on the applied voltage with the size of the droplet is ----.
When the distance of two electrodes is below approximately 70$\mu$m ,
droplets are sticked to an electrode (adhered).
The diagram indecates that the threshold of applied voltage is 
roughly propotional to the distance between two electrodes.

\section{Discussion}
\indent We propose a model to describe the oscillatory-stationary motion of w/o droplets.
In an eqation of motion at a micrometer scale,
a viscosity term is more dominant than inertia term because the Reynolds number,
$R_{e}$, is rather small; $R_{e}=\rho v d/\eta \sim 10^{-9} \ll 1$,
where $\rho (\sim 10^{3} \rm{kg/m^{3})}$ and $\eta (\sim 10^{3} \rm{Pa s})$ are
the density and viscosity of the mineral oil, respectively, and
$v (\sim 10^{-4} \rm{m/s})$ and $d (\sim 10^{-5} \rm{m})$ are the velocity and 
diameter of the water droplet.
Therefore, an over-damped eqation of motion under the constant electric field, $\bm{E}$, is given by
\begin{eqnarray}
k\dot{\bm{x}} = q\bm{E}+\alpha \nabla \bm{E}^{2}
\end{eqnarray}
where $k (=6\pi \eta d \sim 10^{-7} \rm{kg/s})$ is a coefficient of viscosity resistance,
and $k\dot{\bm{x}}$ represents the viscosity resistance for a moving droplet with diameter $d$ 
and velocity $\dot{\bm{x}}$.
$q\bm{E}$ and $\alpha \nabla \bm{E}^{2}$ indicate an electric force and a dielectric force acting
on the droplet with charge $q$ and polarizability $\alpha$  (具体的な値)\cite{jones}.\par
Here we assume that the time-dependent rate od the charge, $q$, is described as 
\begin{eqnarray}
\dot{q} = -\beta \epsilon x^{3} - \frac{q}{t_{0}}
\end{eqnarray}
where $\beta$ is the proportionality coefficient, and $\epsilon$ is the constrant
is proportional to the magnitude of the electrical field.
The first term of this means the time-dependence of charge is in proportion to 
the number of lines of electric force.
The second term means the charge leak, and $t_{0}$ is the relaxing time.\par
We would like to consider the condition where the droplet stays on the same position between two electrodes.
When the electric field can be written as $\bm{E}=(E_{x}, E_{y})$,
The force on the second term in the right hand of Eq. (1) is caused by 
the number of the lines of electric force penetrating the droplet.
Comparing with the size of droplet, the change of $E_{x}$ along $x$ axis can be neglected.  
By considering the symmetry of the system, we simply adapt that the change of $E_{y}^{2}$ along $x$ axis is written as
\begin{eqnarray}
E_{y}^{2} = \epsilon \left( -(x+1)^{2}(x-1)^{2} +2\right).
\end{eqnarray}
Then the $x$ component of eq.(1) is given as
\begin{eqnarray}
k\dot{x} =& qE_{x} + \alpha \partial_{x} E^{2}\nonumber \\
\cong& qE_{x} + \alpha \partial_{x} E_{y}^{2} \nonumber \\
=& qE_{x} - 4\alpha \epsilon x(x+1)(x-1)\nonumber \\
\dot{x} =& \frac{E_{x}}{k}q - \frac{4\alpha \epsilon }{k}x(x+1)(x-1) 
\end{eqnarray}
For simplicity, we introduce the following parameters: 
$E_{x}/k = a$, $4\alpha \epsilon /k = e$ and $ke\beta/4\alpha = \gamma$.
Then Eq.(2) and Eq.(4) can be written as
\begin{eqnarray}
\dot{x} =& -ex(x+1)(x-1) + \alpha q \\
\dot{q} =& -\gamma e x^{3} - q/t_{0}
\end{eqnarray}

\begin{figure}[htbp]
\begin{center}
\includegraphics[]{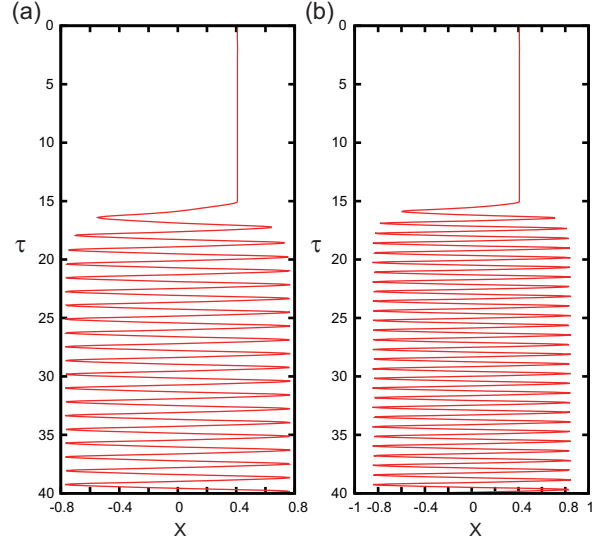}
\end{center}
\caption{Numerical results on the Spatio-Temporal diagram with eqs. (5) and (6), the common parameters are $t_{0} = 0.5$ and $\gamma = 0.1$. The parameters $a$ and $e$ are changed 
at $T=15$. Initial state is the same in both graphs; $a=100$ and $e=2$. After $T=15$ in (a), $a=200$ and $e=4$. 
In (b), $a=250$ and $e=5$. \label{fig:four}}
\end{figure}

FIG.\ref{fig:four} shows the result of numeric calculation with these equations,
where the time and space scales ,$T$ and $X$, are arbitrary.
The change of the distance between the electorodes corresponds to
the change of the magnitude of the electric field.
For example, if $L$ becomes larger, $a$ and $e$ become smaller.
In FIG.\ref{fig:four}, $a$ and $e$ in (b) are larger than those in (a).
The frequency of the back-and-force motion of the droplet is faster 
when the electric field between the electrodes becomes stronger.
Thus, our numerical model reproduces the essential aspect of
the rhythmic motion of a droplet under DC voltage.

\begin{figure}[htbp]
\begin{center}
\includegraphics[]{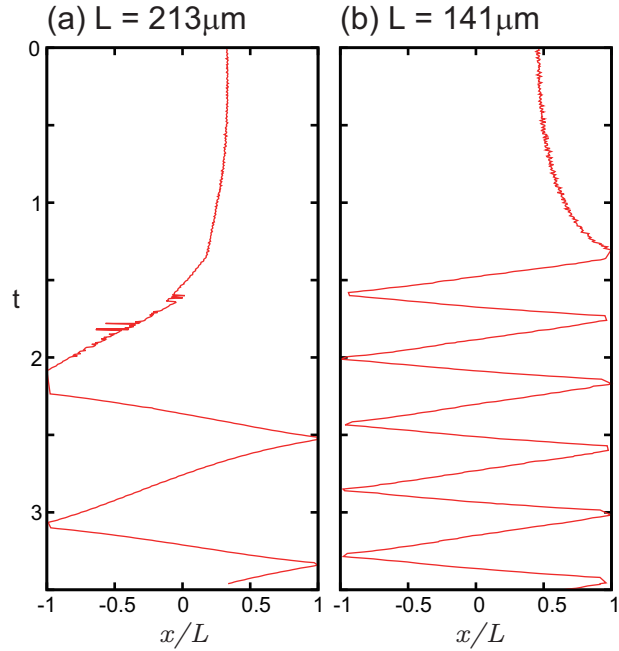}
\end{center}
\caption{Experimental results on the Spatio-Temporal diagram \label{fig:five}}
\end{figure}


\section{Acknowledgements}


\begin{thebibliography}{99}\label{sec:TeXbooks}%
  \bibitem{hiratsuka}Y. Hiratsuka, M. Miyata, T. Tada, and T. Q. P. Uyeda, Proc. Natl. Acad.
Sci. U.S.A. 103, 13618 (2006).
  \bibitem{vandenheuvel}M. G. L. van den Heuvel and C. Dekker, Science 317, 333 (2007).
  \bibitem{hase} Hase, et al., PRE 74, 046301(2006).
  \bibitem{takinoue}Takinoue, et al. Appl. Phys. Lett. 96, 104105 (2010)
  \bibitem{mochizuki}T. Mochizuki, Y. Mori, and N. Kaji, AIChE J. 36, 1039 (1990).
  \bibitem{jung}Y. Jung, H. Oh, and I. Kang, J. Colloid Interface Sci. 322, 617 (2008). 
  \bibitem{ristenpart}W. D. Ristenpart, J. C. Bird, A. Belmonte, F. Dollar, and H. A. Stone, Nature (London) 461, 377 (2009).
  \bibitem{Eow} J. S. Eow, M. Ghadiri, and A. Sharif, Colloids Surf., A 225, 193 (2003).
  \bibitem{teh}S. Teh, R. Lin, L. Hung, and A. Lee, Lab Chip 8, 198 (2008).
  \bibitem{pietrini}A. V. Pietrini and P. L. Luisi, ChemBioChem 5, 1055 (2004).
  \bibitem{tawfik}D. S. Tawfik and A. D. Griffiths, Nat. Biotechnol. 16, 652 (1998).
  \bibitem{hase2}M. Hase and K. Yoshikawa, J. Chem. Phys. 124, 104903 (2006).
  \bibitem{katsura}S. Katsura, A. Yamaguchi, H. Inami, S. Matsuura, K. Hirano, and A. Mizuno, Electrophoresis 22, 289 (2001).
  \bibitem{link}D. R. Link, E. Grasland-Mongrain, A. Duri, F. Sarrazin, Z. Cheng, G. Cristobal, M. Marquez, and D. A. Weitz, Angew. Chem. Int. Ed. 45, 2556 (2006).
  \bibitem{atencia}J. Atencia and D. J. Beebe, Nature (London) 437, 648 (2005).
  \bibitem{voldman}J. Voldman, Annu. Rev. Biomed. Eng. 8, 425 (2006).
  \bibitem{jones}T. B. Jones, Electromechanics of Particles (Cambridge University Press, New York, 1995).  
  
  
  \end{thebibliography}
\end{document}